\documentclass[aps,prl,twocolumn,floatfix,nofootinbib,showpacs,superscriptaddress]{revtex4-1}
\usepackage{amsmath,amsfonts,amssymb,bm}
\usepackage{graphicx}
\usepackage{subfigure}
\usepackage{color}
\usepackage{multirow}%for tabular

\definecolor{purple}{rgb}{0.5,0,0.5}
\definecolor{blue}{rgb}{0.0,0,0.9}
\usepackage[colorlinks=true, pdfstartview=FitV, linkcolor=purple, citecolor= purple, urlcolor=blue]{hyperref}

\begin{document}

%\begin{CJK*}{GBK}{song}

\title{Chemical Freeze-out Parameters via a Non-perturbative QCD Approach}

\author{Jing Chen }
\affiliation{Department of Physics and State
Key Laboratory of Nuclear Physics and Technology, Peking
University, Beijing 100871, China}
\affiliation{Collaborative Innovation Center of Quantum Matter, Beijing 100871, China}

\author{Fei Gao }
\affiliation{Department of Physics and State
Key Laboratory of Nuclear Physics and Technology, Peking
University, Beijing 100871, China}
\affiliation{Collaborative Innovation Center of Quantum Matter, Beijing 100871, China}

\author{Yu-xin Liu }
\email[Corresponding author: ]{yxliu@pku.edu.cn}
\affiliation{Department of Physics and State Key Laboratory of
Nuclear Physics and Technology, Peking University, Beijing 100871,
China}
\affiliation{Collaborative Innovation Center of Quantum Matter, Beijing 100871, China}
\affiliation{Center for High Energy Physics, Peking
University, Beijing 100871, China}

\begin{abstract}
By analyzing the calculated baryon number susceptibility ratios ${\chi_{1}^{B}}/{\chi_{2}^{B}}$ and ${\chi_{3}^{B}}/{\chi_{1}^{B}}$ in two-flavor system via the Dyson-Schwinger equation approach of QCD, we determine the chemical freeze-out temperature and baryon chemical potential in cases of both thermodynamic limit and finite size.
We calculate the center-of-mass energy dependence of the ${\chi_{4}^{B}}/{\chi_{2}^{B}}\, (\kappa \sigma^{2})$ at  the freeze-out line and find an excellent agreement with experimental data
in $\sqrt{S_{NN}^{}} \geq 19.6\,$GeV region when taking into account the finite size effect.
Our calculations indicate that the $\kappa \sigma^{2}$ exhibits a non-monotonic behavior in lower collision energy region.
%
%may take its minimum as $\sqrt{S_{NN}^{}}$ is around even smaller than $10\,$GeV then increase %drastically when $\sqrt{S_{NN}^{}}$ decreases further.
%
%quite large value at $\sqrt{S_{NN}}=5.5\,$GeV.
%
\end{abstract}


\maketitle

%%%%%
\noindent{\bf \it Introduction.}---
Phase transitions of strong interaction matter have been explored for more than forty years since the research may reveal the nature of the early universe matter evolution~\cite{Wambach:2009Review,Fukushima:2011Review,Owe:2013Review,Gupta:2011}.
The transitions include chiral phase transition (from dynamical chiral symmetry to dynamical chiral symmetry breaking) which generates more than $98\%$ of the mass of visible matter
and the confinement transition (hadronization) which slaves the quarks and gluons to hadrons.
They are driven by the temperature ($T$) and the baryon density ($\rho_{B}^{}$) or chemical potential ($\mu_{B}^{}$).
Since the strong interaction can be well described by Quantum Chromodynamics (QCD),
the above mentioned phase transitions are usually referred to as QCD phase transitions.
Moreover, many calculations (see,  {\it e.g.},  Refs.~\cite{Aoki:2006,Aoki:2009,Ejiri:2008,Gupta:2011,Owe:2013Review,Qin:2011a,Xin:2014a,Welzbacher:2014,Liu:2015,Ratti:2006,Schaefer:2007,Fu:2008a,Fukushima:2008,Jiang:2013,Xin:2014b}) have shown that the chiral phase transition at low chemical potential is a crossover at physical quark mass.
Theoretical calculations (see,  {\it e.g.},  Refs.~\cite{Ratti:2006,Schaefer:2007,Fu:2008a,Ejiri:2008,Gupta:2011,Fukushima:2008,Qin:2011a,Xin:2014a,Welzbacher:2014,Liu:2015,Jiang:2013,Xin:2014b}) also indicate that the chiral phase transition at high chemical potential is first order.
Therefore, there would exist a critical end-point (CEP) in the $T$--$\mu_{B}^{}$ plane
at which the first order phase transition turns to crossover.
The position of the CEP or even its existence becomes thus one of the most significant topic in both theories and experiments.
Besides the efforts in theories,
the Beam Energy Scan (BES) program at RHIC, the FAIR at GSI and the NICA at DUBNA  all take the search of the CEP as their investigation focus (see, {\it e.g.}, Refs.~\cite{Melkumov:2012PAN,Odyniec:2012PAN}) and some meaningful information has been provided by the RHIC experiments~\cite{Adamczyk:2014,Luo:2014}.

In experiments, one can measure only the states after the hadronization but not the phase transition directly, and thus the chemical freeze-out line which is defined as the set of states ceasing the inelastic collision of the newly formed hadrons plays the essential role.
Especially, as the chemical freeze-out line  approaches to the CEP, nonmonotonic behavior of conserved charge fluctuations could be observed~\cite{Stephanov:1998,Stephanov:1999,Hatta:2003a,Xin:2014a}.
The freeze--out temperature and chemical potential have then been studied in statistical hadronization model (SHM)~\cite{Becattini:2006,Andronic:2006,Das:2014a,Das:2014}, hadron resonance gas (HRG) model~\cite{Karsch:2011,Alba:2014},
lattice QCD simulations~\cite{Gavai:2011,Bazavov:2012,Borsanyi:2013,Borsanyi:2014} and other models~\cite{Cleymans:2006,Chen:2015}.
In fact, the matter system generated in relativistic heavy ion collision (RHIC) experiment has a finite size and cools in a finite time~\cite{Stephanov:1999,Berdnikov:2000,Abelev:2014}.
The finite size and finite time prevent the correlation length $\xi$ from diverging near the CEP, and smoothen the fluctuations~\cite{Berdnikov:2000}.
Model calculations have shown that the finite size influences both the phase diagram and the thermodynamical properties  drastically~\cite{Palhares:2010,Shao:2006,Bhattacharyya:2013,Bhattacharyya:2015,Bhattacharyya:2015b},  the surface of the system may also play the role~\cite{Berger:1987,Deutsch:1979,Elze:1986,Ke:2014}.
The effects of the finite size and the surface on the chemical freeze-out parameters  will then complement the information for searching the CEP in experiments.
However, different models give contradictory results.
It is therefore imperative to investigate the finite size and the surface effects on the chemical freeze-out parameters with sophisticated QCD  approaches.

It has been known that Dyson-Schwinger equations (DSE), a nonperturbative approach of QCD~\cite{Roberts:DSEinitial,Alkofer:2001Infrared,Roberts:2012Review,Cloet:2014,Chang:2011,Qin:2011b}, have been successful in describing  QCD phase transitions (see, {\it e.g.}, Refs.~\cite{Roberts:2012Review,Qin:2011a,Xin:2014a,Qin:2011PRD,Fischer:2009PRL,Welzbacher:2014,Liu:2015,Bashir:2014,Zong:2014,Wang:2013PRD}) and hadron properties (For recent reviews, see Refs.~\cite{Roberts:2012Review,Cloet:2014}).
We then, in this Letter, take the DSE approach to investigate the chemical freeze-out parameters with  the finite size and surface effects being taken into account. We calculate the baryon number susceptibilities in two light-flavor quark system. By comparing the obtained  baryon number susceptibility ratios ${\chi_{1}^{B}}/{\chi_{2}^{B}}$ and ${\chi_{3}^{B}}/{\chi_{1}^{B}}$ with the experimental data of the net-proton distribution comulant ratios $C_1/C_2$ and $C_3/C_1$ at different collision energies, we determine the freeze-out parameters.
We observe that with the finite size and surface effects being included, the calculated collision energy dependence of the ${\chi_{4}^{B}}/{\chi_{2}^{B}}$ agrees with the experimental data
in $\sqrt{S_{NN}^{}} \geq 19.6\,$GeV region excellently.
Moreover, we propose a nonmonotonic behavior of $\kappa \sigma^2$ in lower collision energy region.
%
%value of ${\chi_{4}^{B}}/{\chi_{2}^{B}}$ at $\sqrt{S_{NN}} = 5.5\;$GeV.
%

\noindent{\bf \it Theoretical Framework.}---
Experimental observations indicate that the yields of pion and proton are much larger than that of kaon~\cite{Das:2014}, we can then simplify the matter generated in RHIC experiments as that including mainly two light flavor quarks. In the system of $u$ and $d$ quarks, baryon number density $n_{B}^{}$ and electric charge density $n_{Q}^{}$ can be fixed with quark number density $n_{u,d}^{}$ as:
\begin{equation}\label{eq:baryonnumber}
 n_{B}^{} = \frac{1}{3}n_{u}^{} + \frac{1}{3}n_{d}^{} \, , \qquad
 n_{Q}^{} = \frac{2}{3}n_{u}^{} - \frac{1}{3}n_{d}^{} \, .
\end{equation}
From Eq.~(\ref{eq:baryonnumber}) one can notice that, if we only consider the baryon number,
the $u$ and $d$ quarks are in exact isospin symmetry.
In this sense, both the $u$ quark and $d$ quark hold the same quark chemical potential $\mu_{q}^{} =\mu_{B}^{}/3$, and the quark number density $n_{q}^{} = 3n_{B}^{}$.

In view of statistical physics, the quark number density can be determined as
\begin{eqnarray}\label{eq:qnumberdistribution}
 n_{q}^{}(\mu_q,T) & = & 2 N_{c} N_{f} Z_{2} \! \int_{-\infty}^\infty \! \frac{d^3 \vec{p}}{(2\pi)^3} f_1(|\vec{p}|; \mu_{q}^{}, T)\, , \quad  \\  \label{eq:fermiondistribution}
 f_{1}^{}(|\vec{p}|; \mu_{q}^{}, T) & = & \frac{T}{2} \! \sum_{j=-\infty}^\infty \!\! tr_{D}^{}(-\gamma_{4}^{} S(i\tilde{\omega}_{j}^{}, \vec{p})),
\end{eqnarray}
where $Z_{2}$ is the quark wave-function renormalization constant, $N_{c} = 3$ the color number,
and $N_{f} = 2$ the flavor number. The $S(i\tilde{\omega}_{j}^{},\vec{p})$ is the quark propagator
which can be fixed by solving the Dyson-Schwinger equation
\begin{eqnarray}\nonumber
&\hspace{-0.2cm} S^{-1}(i \tilde{\omega}_{j}^{}, \vec{p}) = i \vec{\gamma}\cdot\vec{p} +i {\gamma_{4}^{}} \tilde{\omega}_{j}^{} + m_{0}^{} +\frac{4}{3}T\!\! {\sum\limits_{l=-\infty}^{\infty} }  \! \int_{-\infty}^\infty \! \frac{d^3\vec{q}}{(2\pi)^3} g^2\\\label{eq:gapeq}
&\hspace{-0.5cm} \times \! D_{\mu\nu}^{}(\vec{k},\Omega_{jl}^{};T,\mu_{q}^{}) \gamma_{\mu}^{} S(i\tilde{\omega}_{l}^{},\vec{q}) \Gamma_{\nu}^{}(\vec{p},\tilde{\omega}_{j}^{},\vec{q},\tilde{\omega}_{l}^{};T,\mu_{q}^{}) \, ,
\end{eqnarray}
where $m_{0}^{}$ is the current quark mass, $D_{\mu\nu}^{}$ is the dressed-gluon propagator,
$\Gamma_{\nu}^{}$ is the dressed quark-gluon vertex,
$\tilde{\omega}_{j}^{} = \omega_{j}^{} + i \mu_{q}^{}$ and  $\Omega_{jl}^{} = \omega_{j}^{} - \omega_{l}^{}$ with $\omega_{j}^{} = (2j+1)\pi T $, the Matsubara frequency.
In practical calculation we adopt at first stage the rainbow approximation for the vertex $\Gamma_{\nu}^{}(\vec{p},\tilde{\omega}_{m}^{},\vec{q},\tilde{\omega}_{l}^{};T,\mu_{q}^{} ) = \gamma_{\nu}^{}$,
the infrared constant model (QC model)~\cite{Qin:2011b,Xin:2014a,Liu:2015} for the dressed-gluon propagator, $m_{0}^{}\!=\!3.4\,$MeV and $\omega\!=\!0.5\,$MeV as our parameters.

The $k$th order quark number density susceptibility (fluctuation) is obtained as
\begin{equation}
 \chi_{k}^{q} = \frac{1}{\beta^{(k-1)}} \frac{\partial^{(k-1)} {n_{q}^{}}}{\partial^{(k-1)} {\mu_{q}^{}}} \, ,
\end{equation}
where $\beta = 1/T$ and $k = 2,3,4,\cdots$.
The susceptibilities are related to the moments of the multiplicity distributions of the corresponding conserved charges as
\begin{eqnarray}
 \nonumber
\frac{\chi_{1}^{}}{\chi_{2}^{}} = M/\sigma^{2} \, ,  \quad &
&\quad \frac{\chi_{3}^{}}{\chi_{1}^{}} = S\sigma^{3}/M \, ,   \\
\frac{ \chi_{3}^{}}{\chi_{2}^{}} = S\sigma \, , \qquad &
&\quad \frac{\chi_{4}^{}}{\chi_{2}^{}} = \kappa \sigma^{2} \, ,
\end{eqnarray}
where $M$, $\sigma^2$, $S$ and $\kappa$ are correspondingly the mean, the variance, the skewness and the kurtosis of the multiplicity distribution.
By comparing the theoretical net-baryon number fluctuations in terms of temperature and chemical potential with the experimental data one can  determine the freeze--out parameters~\cite{Borsanyi:2014}.

However, the system created in RHIC exists in finite size but not at the thermodynamical limit.
To determine the freeze--out parameters in experiment one has to take the finite size and the surface effects into account.
Assuming the size scale of the systems as $L$, and adopting the anti-periodic condition,
the momentum of a fermion should be $p_{j}^{} = (2j+1)\pi/L$.
The finite size effect can thus be roughly incorporated by a non-zero momentum cut-off~\cite{Bhattacharyya:2013,Bhattacharyya:2015}
$|p|_{min}^{} = \pi/L$.
It corresponds to an infrared momentum cut-off in Eqs.~(\ref{eq:qnumberdistribution}) and (\ref{eq:gapeq}).
It is remarkable that such an $L$ is not the size of the fireball, but an effective scale that the ingredients of the matter can interact.

We also incorporate the effect of the surface through the multiple reflection expansion (MRE) approximation.
In the MRE approximation, the thermodynamical quantities of a droplet composed of quarks can be derived from a density of states in the form~\cite{Berger:1987,Deutsch:1979,Elze:1986}
\begin{equation}\label{eq:MRE}
 \frac{dN}{dp} = 6\left[ \frac{p^{2} V}{2 \pi^{2}} + f_{s}^{}(\frac{p}{M}) p S
 + f_{c}^{} (\frac{p}{M}) C + \cdots \right] \, ,
\end{equation}
where $V$ is the volume of the droplet, $S=4\pi L^2$ and $C=8\pi L$ are the area, the extrinsic curvature of the surface of the droplet, respectively.
The $f_{s}^{}(\frac{p}{M})$ and $f_{c}^{} (\frac{p}{M})$ are the contributions to the density of states 
from the surface and the curvature, respectively, with $p$ being the momentum and $M$  the
constituent quark mass.
Rigorously, one should solve the coupled equations~\cite{Shao:2006}.
For simplicity we set $M$ in Eq.~(\ref{eq:MRE}) to be that calculated from a system of size $L$.

\begin{figure}[htb]\vspace*{-0.4cm}
  \includegraphics[width=0.40\textwidth]{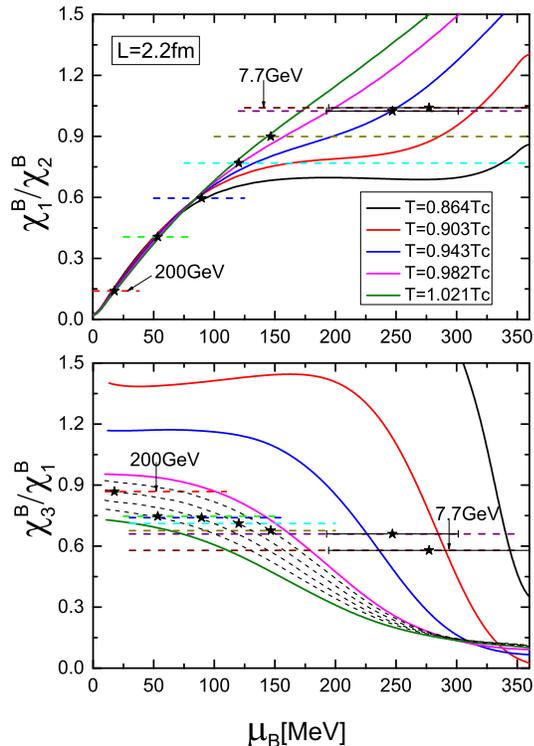}
\vspace*{-0.4cm}\caption{\label{fig:x123r2-2} (color online) Calculated baryon chemical potential dependence of the fluctuation ratios  ${\chi_{1}^{B}}/{\chi_{2}^{B}}$ (upper panel)
and ${\chi_{3}^{B}}/{\chi_{1}^{B}}$ (lower panel)
of the system with $L=2.2\,$fm at several values of temperature near the $T_{c}$.
The dashed horizontal lines stand for the experimental values of the efficiency-corrected
$C_{1}/C_{2} = M/\sigma^2$ and $C_{3}/C_{1} = S\sigma^3/M$ of net-proton multiplicity distributions in the central collisions at $\sqrt{S_{NN}^{}} = 200$, $62.4$, $39$, $27$, $19.6$, $11.5$, $7.7\,$GeV given in Ref.~\cite{Luo:2014}.
The stars label our assigned freeze-out points.}
\end{figure}

%%%%% Results
\noindent{\bf \it Freeze-out Parameters.}---
We have carried out calculations with $L=\infty$(the thermodynamical limit) and various finite values of $L$.
The calculations manifest that the fluctuations (skewness, kurtosis, etc.) in the $T$--$\mu_{B}^{}$ plane behave qualitatively the same as those given in Ref.~\cite{Xin:2014a}, respectively,  except for the pseudo-critical temperature $T_{c} $. The $T_{c}^{}$ determined by the chiral susceptibility criterion is $151.3\,$MeV in case of $L = \infty$, $127.3\,$MeV when $L=2.2\,$fm and $129.4\,$MeV for  $L=2.2\,$fm with the surface effect being included via the MRE correction.
The obtained $\mu_{B}^{}$ dependence of the baryon number susceptibility ratios  ${\chi_{1}^{B}}/{\chi_{2}^{B}}$
and ${\chi_{3}^{B}}/{\chi_{1}^{B}}$ in case of $L=2.2\,$fm at several values of temperature
are shown in Fig.~\ref{fig:x123r2-2}.
It is evident that our results agree with the lattice QCD results~\cite{Borsanyi:2014} qualitatively very well.
In order to extract the freeze-out parameters, we plot the experiment values of the efficiency-corrected cumulant ratios $C_{1}/C_{2} = M/ \sigma^2$
and $C_{3}/C_{1} = S\sigma^3/M$ of net-proton multiplicity distributions
in central collisions~\cite{Luo:2014} as horizontal lines.
By fitting our calculated ${\chi_{1}^{B}}/{\chi_{2}^{B}}$ and ${\chi_{3}^{B}}/{\chi_{1}^{B}}$ values in terms of $T$ and $\mu_{B}^{}$ with the experimental data we get the freeze--out parameters $( \mu_{B}^{f} , T^{f} )$.
The obtained results when $L= \infty$ (only available for $\sqrt{S_{NN}^{}} \! \geq \! 27\,$GeV), $L=2.2\,$fm and $L=2.2\,$fm with the MRE correction  are listed in Table~\ref{tab:freezeout}.

\begin{table}[t!]
\caption{\label{tab:freezeout} Calculated freeze-out parameters $( \mu_{B}^{f}, T^{f} )$ in case of $L= \infty$, $L=2.2\,$fm and $L=2.2\,$fm with the MRE correction ($T_{c}$, $T^{f}$ and $\mu_{B}^{f}$ are in unit MeV and $\sqrt{S_{NN}^{}}$ in GeV).}
 \begin{tabular}{|c|c|c|c|c|c|c|}
\hline
\multirow{3}{*}{~~$\sqrt{S_{NN}}$~~} & \multicolumn{2}{|c|}{~~$L=\textrm{infinity}$~~} & \multicolumn{2}{|c|}{~~$L=2.2\,$fm~~} & \multicolumn{2}{|c|}{~$L=2.2\,$fm+MRE}\\
     & \multicolumn{2}{|c|}{($T_{c}=151.3$)} & \multicolumn{2}{|c|}{($T_{c}=127.3$)} & \multicolumn{2}{|c|}{($T_{c}=129.4$)} \\
\cline{2-7}
     & $\mu_{B}^{f}$ &   $T^{f}$   & $\mu_{B}^{f}$ &   $T^{f}$   &$\mu_{B}^{f}$ &   $T^{f}$   \\
\hline
200  &  23.1  & 157.5 &   17.6  & 127.0 & 18.3  & 128.5 \\
\hline
62.4 &  67.3  & 157.5 &   53.4  & 128.9 & 55.3  & 130.2 \\
\hline
39   &  105.8 & 155.5 &   89.6  & 127.9 & 87.1  & 129.3 \\
\hline
27   &  173.8 & 149.5 &   120.2 & 127.0 & 123.0 & 128.2 \\
     & $\pm$ 35.6 & $\pm$ 4.0 &       &       &       &       \\
\hline
19.6 &   ---   &  ---   &   146.6 & 126.0 & 155.7 & 126.7 \\
\hline
11.5 &   ---   &  ---   &  256.1  & 119.5 & 287.2 & 118.0 \\
     &        &       &  $\pm$ 45.1  & $\pm$ 4.0 & $\pm$ 66.1 &  $\pm$ 5.5  \\
\hline
7.7  &   ---   &  ---   &  283.5  & 118.0 & 277.9 & 119.5 \\
     &        &       &  $\pm$ 76.5  & $\pm$ 6.0 & $\pm$ 76.1 &  $\pm$ 7.0  \\
\hline
\end{tabular}
\end{table}

\begin{figure}[htb]\vspace*{-0.2cm}
\includegraphics[width=0.40\textwidth]{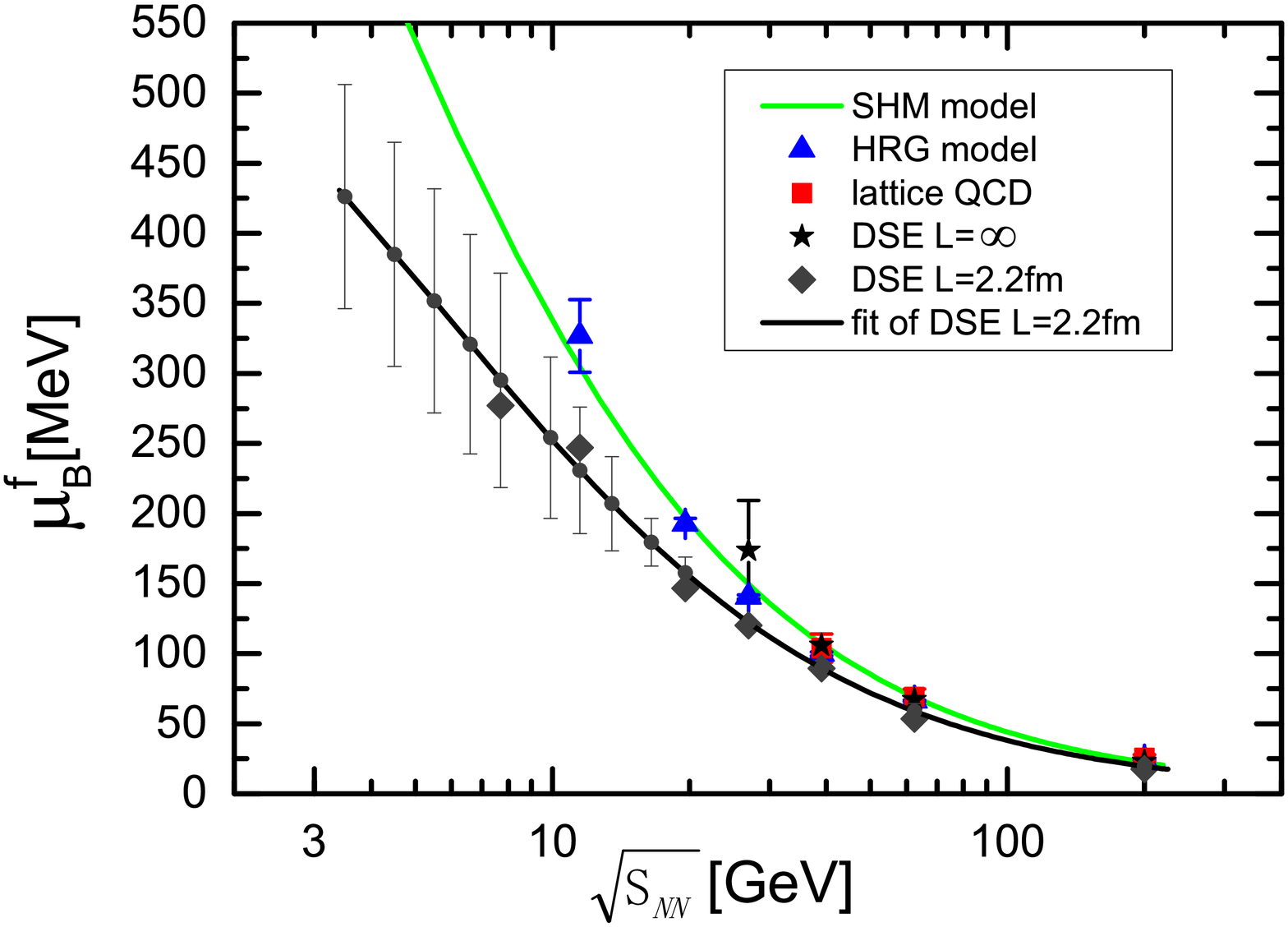}
\vspace*{-0.2cm}\caption{\label{fig:mube} (color online) Comparison of presently obtained $\sqrt{S_{NN}^{}}$ dependence of the baryon chemical potential in cases of $L = \infty$ and $L=2.2\,$fm with those given in lattice QCD simulation~\cite{Borsanyi:2014}, HRG model~\cite{Alba:2014}
and the parameterized one in SHM model~\cite{Andronic:2006}. The error bars on our fitted line in the $L=2.2\,$fm case label the uncertainty of the freeze-out chemical potential.}
\end{figure}

 We illustrate the presently calculated relation between the baryon chemical potential $\mu_{B}^{f}$ and the center-of-mass energy of the collision, $\sqrt{S_{NN}^{}}$, and the comparison with those given in lattice QCD simulations ({\it e.g.}, Ref.~\cite{Borsanyi:2014}) and model calculations ({\it e.g.}, Refs.~\cite{Alba:2014,Andronic:2006}) in Fig.~\ref{fig:mube}.
We see from Fig.~\ref{fig:mube} that our freeze-out baryon chemical potential in case of $L= \infty$  and that when $L=2.2\,$fm match the lattice QCD result and model calculation results very well in the region $\mu_{B}^{} < 110\,$MeV,
while those in case of $L=2.2\,$fm deviate from previous results in the $\mu_{B}^{} > 110\,$MeV range.
We then re-fit our freeze-out conditions in case of $L=2.2\,$fm as:
\begin{eqnarray}
\label{eq:mapmuBf}
 \mu_{B}^{f} & = & \frac{c}{1+d \sqrt{S_{NN}^{}} } \, ,  \\     \label{eq:mapTf}
 T^{f} &  = & T^{0} \Big[ 1 - a \Big( \frac{\mu_{B}^{f}}{T^{0}} \Big)^2
 - b \Big( \frac{\mu_{B}^{f}}{T^{0}} \Big)^4 \Big] \, .
\end{eqnarray}
The obtained parameters are $T^{0} = 128.3\,$MeV, $a = 0.0122$, $b = 0.000990$, $c = 676.8\,$MeV,
and $d = 0.168\,\textrm{GeV}^{-1}$ for the case of $L=2.2\,$fm without MRE correction.
The fitted curve $\mu_{B}^{f}(\sqrt{S_{NN}^{}})$ is also displayed in Fig.~\ref{fig:mube}.
With these formulas  one can have the freeze--out parameters $({\mu_{B}^{f} }, {T^{f}})$
of the system generated in any collision energy.
For example, corresponding to $\sqrt{S_{NN}^{}}=5.5\,$GeV, $14.5\,$GeV,
$( \mu_{B}^{f}, T^{f} ) = ( 351.8, 109.5)\,\textrm{MeV}$, $( 197.0, 123.9)\,\textrm{MeV}$, respectively.

\begin{figure}[htb]\vspace*{-0.5cm}
  \includegraphics[width=0.40\textwidth]{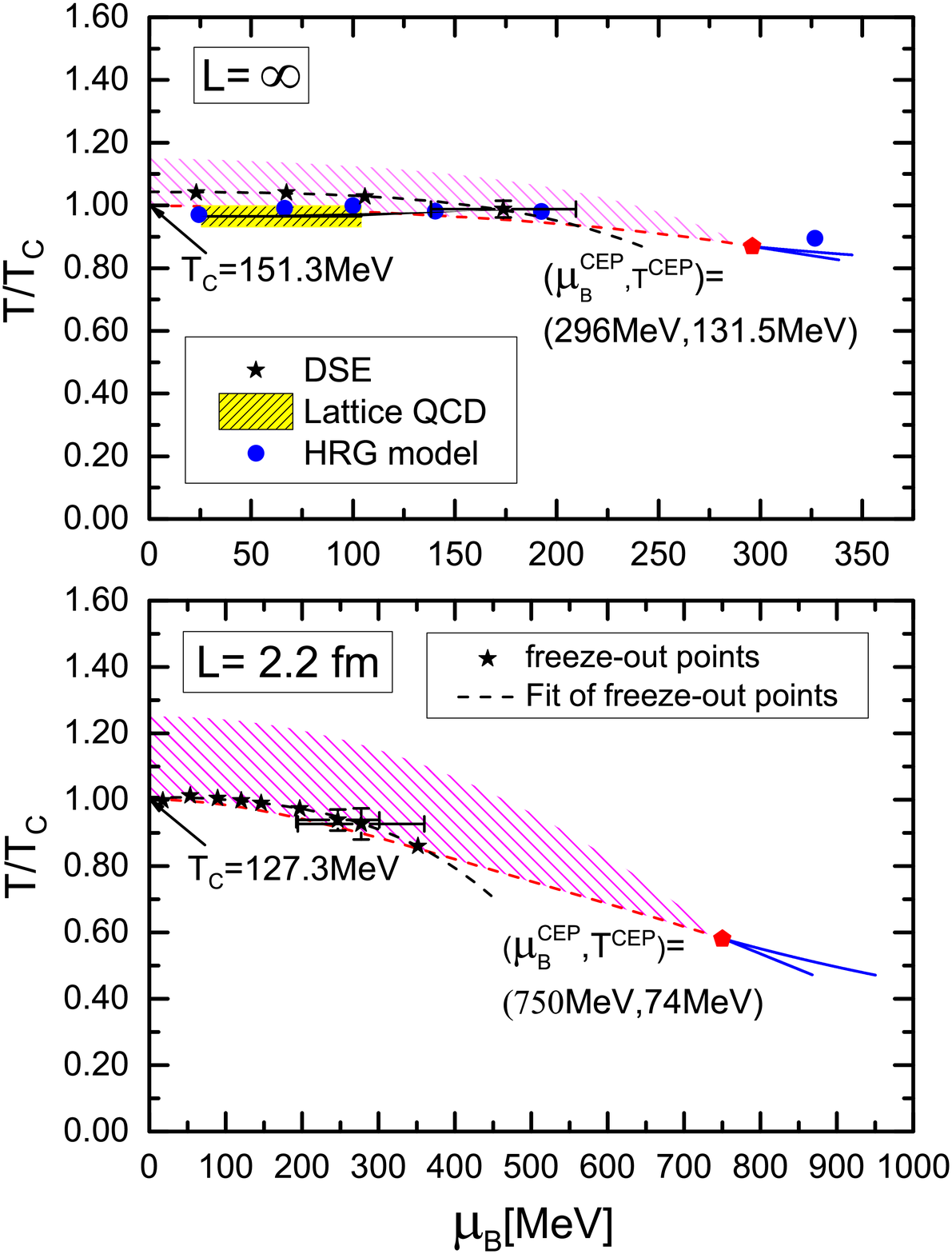}
\vspace*{-0.4cm}\caption{\label{fig:phase} (color online) Calculated QCD phase diagram in case of thermodynamic limit (upper panel) and that in case of $L= 2.2\,$fm (lower panel), together with the calculated freeze-out lines (in case of $L = \infty$, we show also those given in lattice QCD~\cite{Borsanyi:2014} and HRG model~\cite{Alba:2014} for comparison). }
\end{figure}

%%%%% Results
\noindent{\bf \it Phase Diagram and Further Prediction.}---
With the quark propagator obtained by solving the DSE, we can get the temperature and chemical potential dependence of the quark condensate and the dynamical mass of the quark, which are commonly regarded as appropriate order parameters of chiral phase transition.
Taking the chiral susceptibility criterion~\cite{Qin:2011a} we determine the lower boundary of the chiral phase crossover region. The obtained pseudo-critical temperatures at zero chemical potential in the two cases are listed in Table~\ref{tab:freezeout}.
As for the upper boundary, we assign it as the set of the states for the dynamical quark mass at zero momentum to decrease to the $10\%$ of that at $T\! = \!0$ and $\mu_{B}^{} \! = \! 0$.
The obtained crossover regions in cases of $L\! =\! \infty$ and $L\! = \! 2.2\,$fm are shown as the shadowed regions in Fig.~\ref{fig:phase}.
With the chiral susceptibility criterion~\cite{Qin:2011a} or the fluctuation criterion~\cite{Xin:2014a},
we determine the boundaries of the first order transition region and the location of the CEP.
The obtained results in the two cases are displayed in Fig.~\ref{fig:phase}.
We illustrate also the presently obtained chemical freeze--out lines in the two cases in Fig.~\ref{fig:phase}. The figure manifests apparently that $T^{f}(\mu_{B}^{} =0) =157.5\,$MeV is a few MeVs higher than the $T_{c} (\mu_{B}^{} =0) =151.3\,$MeV in the thermodynamical limit ({\it i.e.}, with $L=\infty$), but $T^{f} (\mu_{B}^{} =0) \ngtr T_{c}(\mu_{B}^{} = 0)$ in case of $L=2.2\,$fm.
It indicates that the result including the finite size effect is more reasonable.
 We also notice that the finite size effect shifts the location of the CEP to higher baryon chemical potential and lower temperature drastically,
which is consistent with the behavior given in phenomenological model calculations~\cite{Palhares:2010,Bhattacharyya:2013}.

\begin{figure}[htb]\vspace*{-0.2cm}
  \includegraphics[width=0.42\textwidth]{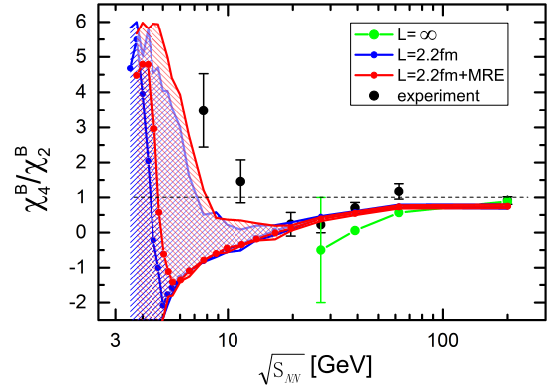}
\vspace*{-0.4cm}\caption{\label{fig:x4x2} (color online) Calculated collision energy $\sqrt{S_{NN}}$ dependence of $\kappa \sigma^2 = {\chi_{4}^{B}}/{\chi_{2}^{B}}$ at the freeze-out line.
The black circles are the experimental values~\cite{Luo:2014}, the green circles stand for our results in case of infinite volume, the blue and red points denote our results in the case of $L=2.2\,$fm without and with MRE correction respectively. The shadowed region(s) displays the uncertainties corresponding to those of the $\mu_{B}^{f}$ and $T^{f}$.}
\end{figure}

It is known that the $\kappa \sigma^2 = {\chi_{4}^{}}/{\chi_{2}^{}}$ is a direct observable in experiment and may demonstrate the property of the states around the CEP well.
We calculate ${\chi_{4}^{B}}/{\chi_{2}^{B}}$ in the $T$--$\mu_{B}^{}$ plane and pick out the value along the freeze-out line to get the $\sqrt{S_{NN}^{}}$ dependence of ${\chi_{4}^{B}}/{\chi_{2}^{B}}$.
The obtained results in case of thermodynamical limit, finite size with $L = 2.2\;$fm
and $L=2.2\,$fm with MRE correction are depicted in Fig.~\ref{fig:x4x2}.
It is apparent that,
without considering the finite size effect,  our calculated  ${\chi_{4}^{B}}/{\chi_{2}^{B}}$ decreases more rapidly than the experimental data as the $\sqrt{S_{NN}^{}}$ descends.
With the finite size effect being taken into account, we can reproduce the experimental data
in $\sqrt{S_{NN}^{}} \geq 19.6\,$GeV region excellently and the MRE correction improves the agreement a little further.
In the lower collision energy region, the $\kappa \sigma^2$ exhibits a non-monotonic behavior which may reach its minimum as $\sqrt{S_{NN}^{}}$ is about even smaller than $10\,$GeV, and then increases drastically when $\sqrt{S_{NN}^{}}$ decreases further.

%%%%% Summary
\noindent{\bf \it Summary.}---
In summary, we calculate the baryon number susceptibilities in a two-flavor quark system via the DSE approach of QCD in case of not only thermodynamic limit but also finite size.
By comparing the calculated ratios ${\chi_{1}^{B}}/{\chi_{2}^{B}}$ and ${\chi_{3}^{B}}/{\chi_{1}^{B}}$
with the experimental data of the net-proton multiplicity distribution in BES at RHIC, we obtained the  temperature and the baryon chemical potential at the chemical freeze-out states.
We calculate also the collision energy dependence of the $\kappa \sigma^{2}$
at the freeze-out line and observed an excellent agreement with experimental data in the region $\sqrt{S_{NN}} \geq 19.6\,$GeV when taking into account the finite size effect.
It shows that the finite size effect is significant in studying the QCD phase transitions with RHICs, while the surface effect offers slight correction further.
The obtained collision energy $\sqrt{S_{NN}^{}}$ dependence of the $\kappa \sigma^{2}$ exhibits a non-monotonic behavior in lower collision energy region.
%

%\bigskip

{\bf \it Acknowledgments:}
The work was supported by the National Natural Science Foundation of China under Grant Nos.\ 10935001, 11175004 and 11435001, and the National Key Basic Research Program of China under Grant Nos.\ G2013CB834400 and 2015CB856900.

%\end{CJK*}

\end{document}